\documentclass[pre,twocolumn,floatfix,superscriptaddress,aps]{revtex4}

\usepackage{graphicx}
\usepackage[usenames]{color}

\begin{document}

\title{{Improved}
effective-range expansions for small {and large} values of scattering length}

\author{ Adhikari S K \footnote{Adhikari@ift.unesp.br;  http://www.ift.unesp.br/users/Adhikari}
} 
\address{
Instituto de F\'{\i}sica Te\'orica, UNESP - Universidade Estadual Paulista, 01.140-070 S\~ao Paulo, S\~ao Paulo, Brazil
}


\begin{abstract}

The textbook   effective-range expansion of scattering theory is useful  in the 
analysis of low-energy
scattering phenomenology 
when the scattering length $|a|$ is much larger than the range $R$  of the scattering 
potential: $|a|\gg R$.  Nevertheless,
the same has been used for systems 
where the  scattering length is much smaller than the 
range of the potential, which could be the case in many scattering problems.  
We suggest     {and  numerically study improved two-parameter effective-range expansions for 
the cases $|a| > R$ and $|a| < R$.  The improved  effective-range  expansion for  $|a| > R$  reduces to the textbook expansion for  $|a|/R \gg 1$. }

\end{abstract}



\maketitle

{\section {Introduction}} 
    
The textbook effective-range expansion of scattering theory with short-range potential in three spatial dimensions is the following Taylor series of the function $k\cot \delta$ analytic in 
$k^2$ \cite{thaler,thaler2, amjp, ahmed, cit1,bethe}
\begin{eqnarray}\label{eq1}
k \cot \delta= -\frac{1}{a}+\frac{1}{2}r_0 k^ 2+ O(k^4)+...,
\end{eqnarray}
where  $k^2$ is the energy in units of $\hbar^2/2m$,  $m$ is the 
reduced mass, $a$ is the scattering length,  $r_0$ is the effective range of interaction, and $\delta $ is the $S$-wave phase shift 
satisfying  the low energy limit $\lim_{k^2\to 0}\delta = -ka$.  
At small energies when the higher order terms in    (\ref{eq1}) can be neglected, the  $k\cot\delta$-$k^2$ plot 
is a straight line for large values of scattering length $|a|$, which can be used to predict the phase shift $\delta$ at different energies. The effective range expansion (\ref{eq1}) was indeed used in the study 
and analysis of low-energy  scattering problems in 
nuclear  \cite{cit1,bethe,cit2} and  atomic \cite{atom}   physics.   In the general case, $r_0$ is just a coefficient of the Taylor series expansion of $k\cot \delta$. However, for a short-range potential, e.g., a square-well, a Gaussian well, a Yukawa well, or an exponential well, etc., it has been shown \cite{cit1,bethe,cit2} that for large values of scattering length $|a|$, the parameter $r_0$   relates approximately to the range of the potential. For $|a|/R\gg 1$, this relation becomes exact for a square-well potential with range $R$.  Similar effective range expansions have been suggested for potential 
scattering in  one \cite{1d}
and two  spatial dimensions \cite{2d}. 

The applicability of the effective-range expansion is limited to the cases where the scattering can be described by a short-range potential, which may not be the case for  scattering by many composite systems. In addition one should have the condition $|a|/R \gg 1$ \cite{thaler}, otherwise the effective range $r_0$ could  be very different from  the actual range $R$ of the potential.

Originally, the effective-range expansion (\ref{eq1}) was applied \cite{bethe,cit2} to the triplet $S$  ($^3$S$_1$) and singlet $S$ ($^1$S$_0$) channels of neutron-proton scattering at low energies, where the scattering lengths were $a\approx 5$ fm and $\approx -17$ fm, respectively, with $|a|$ much larger than the range of the interaction potential ($R \sim 1$ fm, $|a|/R\gg 1$) \cite{bethe,cit2}.     For small values of scattering length  $|a|\to 0$ the function $k\cot\delta$ of   (\ref{eq1}) develops a pole in $k^2$ near $k^2=0$ as  found in the scattering of neutron-deuteron and pion-pion \cite{adhiplb,nd} systems 
 and for $a=0$ this pole appears at $k^2=0$.  Then the first term on the right-hand-side (rhs) of   (\ref{eq1}) diverges and for $k\cot \delta$ to be finite at low energies, the coefficients of the subsequent terms in this equation has to diverge, e. g., $r_0 \to \infty$.  
  Consequently, the $k\cot\delta$-$k^2$ plot  ceases to be  a straight line at small energies and the effective-range function $k\cot\delta$ is not a convenient function for expansion in $k^2$ as $a\to 0$.
Indeed, the effective range expansion (\ref{eq1}) breaks down in this limit. 
The deficiency of the function (\ref{eq1}) for small $a$ ($k|a|, |a|/R\ll 1$) was discussed \cite{thaler} and 
noted   \cite{ahmed,ket,ket2} before.  For small phase shifts $\delta\approx \tan \delta-\tan^3 \delta/3$, which leads to an alternate parametrization of phase shift for $a\to 0$ \cite{ket}:  
\begin{equation}\label{ketterle}
-\frac{\delta}{k}=a-\biggr[\frac{a^3}{3}-\frac{a^2r_0}{2}\biggr]k^2+....
\end{equation}
However, no proper  effective-range expansion was given in \cite{ket} for $|a|/R  \ll 1$.  It has been pointed out \cite{ket2} that in studies of scattering in cold atoms often the scattering length could be tuned to   zero  near a Feshbach resonance \cite{fesh} 
and in such a case the necessity of an
 appropriate effective range expansion in this limit cannot be over-emphasized. In this study,
 we provide {two such two-parameter (scattering length and range)}  effective-range expansions valid in the limit $|a| <R $, viz.   (\ref{eq9}) and (\ref{eq91}),
   suitable for low-energy scattering phenomenology.  Equation (\ref{eq9}) should be considered to be complimentary to (\ref{eq1})
and without (\ref{eq9}) the effective range theory, a fundamental topic in an introductory course  
on quantum scattering theory,  
cannot be considered complete. { Expansion (\ref{eq91})  reduces to  (\ref{eq9}) in the limit $|a| \ll R $. In addition, we provide an improved two-parameter (scattering length and range) effective-range expansion for large values of scattering length: $|a|> R$, viz. 
(\ref{eq811}).}

\vskip  .7 cm

\section{Effective-range expansion}

A convenient  function  for an expansion  for  $|a|/R \ll 1$ is $k^{-1}\tan \delta$ \cite{adhiplb}, which is the {reciprocal} of the function (\ref{eq1}). A Taylor series expansion of the
function  $k^{-1}\tan \delta$  for small  $k^2$ should have the following form 
\begin{eqnarray}\label{eq2}
k^{-1}\tan \delta =-a +b k^2 +O(k^4) +...,
\end{eqnarray}
where $b$ is a coefficient of expansion { with  the dimension of length cubed}.  If we {take the reciprocal of the}  expansion (\ref{eq1}) for   $|a|/R \ll 1$
we obtain \cite{thaler}
\begin{eqnarray}\label{inv}
k^{-1}\tan \delta =-a -\frac{1}{2}a^2 r_0 k^2 +O(k^4)+...
\end{eqnarray}
Expansion  (\ref{inv}) also   follows   from   (\ref{ketterle}).

Although   (\ref{inv}) relates the parameter $b$ of   (\ref{eq2}) to the parameter $r_0$ of   
(\ref{eq1}) through $b=-a^2 r_0/2$,
we would like to give a physical interpretation of the parameter $b$ for $|a|/R \ll 1$ by relating it to the range { and scattering length}  of a finite-range interaction potential.  We recall that the name effective range was attributed to $r_0$ of expansion (\ref{eq1}) after a comparison with the same for a square-well potential with a well-defined range.  

      The function $k^{-1}\tan \delta$ can be related to the $S$-wave scattering solution of a short-range potential $V(r)$. 
If we define $u(r)\equiv r\psi(r)$, where $\psi(r)$ is the  $S$-wave radial wave function for potential $V(r)$, and $r$ the radial coordinate, then $u(r)$ satisfies \cite{bethe}
\begin{eqnarray}  \label{pot}
u''(r)+[k^2-V(r)]u(r)=0,
\end{eqnarray}
where prime denotes radial derivative. We are using units $\hbar=2m=1$. The corresponding free-particle wave function
$v(r)$
 satisfies
\begin{eqnarray}  \label{free}
v''(r)+k^2v(r)=0.
\end{eqnarray}
 We consider the solution 
\begin{equation}\label{vv}
v(r) =\frac{\sin(kr)}{k\cos \delta}
\end{equation}
 of   (\ref{free}) and the solution of   (\ref{pot}) with the asymptotic behavior 
\begin{equation}\label{as}
\lim_{r\to \infty}u(r) =\frac{\sin(kr+\delta)}{k},
\end{equation}
 and with property $u(0)=0$.  
Multiplying   (\ref{free}) by $u(r)$ and   (\ref{pot}) by $v(r)$ and subtracting and integrating over $r$ we get \cite{bethe}
\begin{eqnarray}  \label{diff}
 \int_0^\infty dr \big[v(r)u''(r)-u(r)v''(r) \big]  = \int_0^\infty dr u(r) v(r) V(r).
\end{eqnarray}
Integrating by parts and using the boundary conditions  $u(0)=v(0)=0$ we get
\begin{eqnarray}
\lim_{r\to \infty} [v(r)u'(r)-u(r)v'(r)] = \int_0^\infty dr u(r) v(r) V(r). 
\end{eqnarray}
Using the asymptotic boundary conditions  (\ref{vv}) and (\ref{as})  on $v(r)$ and $u(r)$, respectively,  we get
 \begin{eqnarray}\label{fin}
k^{-1}\tan \delta = - \int_0^\infty dr u(r) v(r) V(r). 
\end{eqnarray}
Equation (\ref{fin}) relates the present expansion function to the solution of the scattering problem. Bethe \cite{bethe}
also related the expansion function (\ref{eq1}) to the solution of the scattering problem.

    To attribute a physical interpretation to the parameter $b$ of   (\ref{eq2}) for  $|a|/R \ll 1$, we consider the analytically known result for $k\cot \delta$ for $S$-wave scattering by an attractive  square-well potential of range $R$ and depth $\beta^2$ \cite{cit3}:
\begin{eqnarray}\label{eq3}
k^{-1} \tan \delta = \frac {k \tan \gamma R -\gamma  \tan k R}{k^2\tan \gamma R \tan k R+\gamma k},
\end{eqnarray}
where $\gamma=\sqrt{\beta^2+k^2}$. {This result can also be obtained by a direct evaluation 
of the integral on the rhs of Eq. (\ref{fin}).}  
The scattering length $a$   for this potential is
\begin{eqnarray}\label{eq4}
a\equiv  -\lim_{k^2\to 0} k^{-1}\tan \delta =R-\frac{\tan \beta R}{\beta}.
\end{eqnarray}
For this square-well potential the scattering wave function of   (\ref{pot}) with the asymptotic behavior 
(\ref{as}), has the following form  for  $r<R$ 
\begin{eqnarray}\label{uu}
u(r)= \frac{\cos(kR+\delta)}{\gamma\cos \gamma R} {\sin \gamma r}.
\end{eqnarray}
The function (\ref{uu}) is obtained after solving   (\ref{pot}) for $r<R$ and matching the wave functions for $r>R$ and $r<R$  and their  derivatives at $r=R$ in usual fashion \cite{cit3}.   
Using   (\ref{vv}) and (\ref{uu}), a low-energy expansion of  the rhs of   (\ref{fin}) leads to   
\begin{eqnarray}\label{xx}
  k^{-1}\tan \delta & =& \biggr(-R + \frac{\tan\beta R}{\beta} \biggr) (1-Rk \tan \delta)  \nonumber \\
 & +&\frac{1}{6\beta^3}\biggr( \beta R(3+4\beta^2 R^2)-3(1+2\beta^2R^2)\tan \beta R  \nonumber \\ 
&+& 3 \beta R \tan^2 \beta R \biggr) k^2 +...
\end{eqnarray}
Now recalling $\lim_{k^2\to 0}\delta =-ka$ and using   (\ref{eq4}),   (\ref{xx}) can be rewritten as 
   \begin{eqnarray}\label{eq6}
\frac{\tan \delta}{k} = -a +\biggr[\frac{a}{2\beta^2}+\frac{R^3}{6}-\frac{a^2 R}{2}\biggr]k^2+O(k^4)+...,
\end{eqnarray}
consistent with   (\ref{eq2}).  A low-energy expansion of the analytic result (\ref{eq3}) also yields   (\ref{eq6}), which is useful  for $|a|/R < 1$.

\begin{figure}[!t]
\begin{center}
\includegraphics[width=\linewidth]{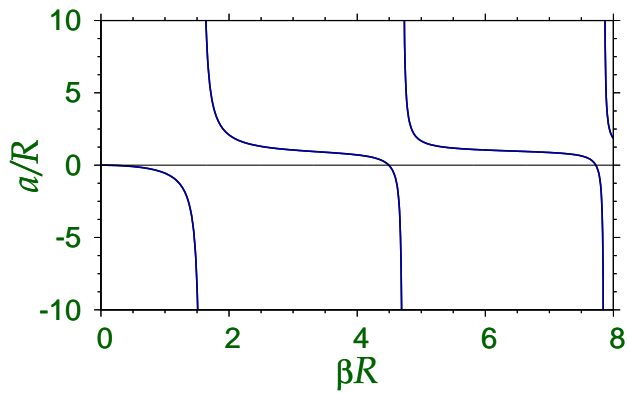} 
\caption{Dimensionless  scattering length $a/R$ versus $\beta R$ plot of Eq, (\ref{eq4}).
   }\label{fig1}
\end{center}

\end{figure}

The analytic expression (\ref{eq3})    yields the  following low-energy $(k^2\to 0)$ Taylor series expansion  \cite{adhiplb}:
 \begin{eqnarray}\label{eq5}
k\cot \delta = -\frac{1}{a}+\biggr[\frac{R}{2}-\frac{R^3}{6a^2}-\frac{1}{2a\beta^2}\biggr]k^2 +O(k^4)+...,
  \end{eqnarray}
useful for  $|a|/R > 1$. Mathematically, expansion (\ref{eq5}) is just the reciprocal of the expansion (\ref{eq6}).
If we impose the  conditions  $|a|/R \ll 1$  and $|a|/R \gg 1$   in 
  (\ref{eq6}) and (\ref{eq5}),  we get  {the lowest-order two-parameter expansions} 
 \begin{eqnarray}\label{eq8}
\frac{\tan \delta}{k} = -a +\frac{1}{6}R^3k^2+O(k^4)+...,\\
\label{eq7}
k\cot \delta = -\frac{1}{a}+\frac{1}{2}Rk^2 +O(k^4)+...,
  \end{eqnarray} 
respectively. {Keeping the next-order terms in   (\ref{eq6}),  we get the improved two-parameter effective-range expansion for $|a|/R <1$:
 \begin{eqnarray}\label{eq81}
\frac{\tan \delta}{k} = -a +\left[\frac{R^3}{6}-\frac{a^2 R}{2}\right]k^2+O(k^4)+...{}.
\end{eqnarray}
{For  $|a|/R <1$, the term $ a/(2 \beta^2)$ in  (\ref{eq6}) is found to be much smaller than 
the other terms and hence is neglected.}
For $|a|/R\gg 1$, Eq. (\ref{eq4}) yields $\beta R = \pi/2$. Using this condition in Eq. (\ref{eq5}) we get the 
improved two-parameter effective-range expansion for $|a|/R >1$:
 \begin{eqnarray}
\label{eq71}
k\cot \delta = -\frac{1}{a}+\left[\frac{R}{2} -\frac{R^3}{6a^2}-\frac{2R^2}{\pi^2 a}\right]k^2 +O(k^4)+...,
  \end{eqnarray}  
Comparing}   (\ref{eq1}) and (\ref{eq7}), we realize  that the effective range $r_0$ is the range of the potential in the limit $|a|/R\gg 1.$ Comparing    (\ref{eq2}) and (\ref{eq8}),  we obtain the following 
effective-range expansion for  $|a|/R \ll  1$ \cite{adhiplb}:
\begin{eqnarray}\label{eq9}
k^{-1}\tan \delta = -a + \frac{1}{6}\tilde r_0^3k^2+ O(k^4)+...,
\end{eqnarray}
with the identification $b=\tilde r_0^3/6$, where we call $\tilde r_0$ an effective range as in   (\ref{eq1}), which becomes the actual 
range $R$ of the potential well as $a\to 0$, viz.   (\ref{eq8}). { A consideration of Eq. (\ref{eq81})
imply the following improved two-parameter effective-range expansion for  $|a|/R < 1$:
\begin{eqnarray}\label{eq91}
k^{-1}\tan \delta = -a +\left[ \frac{\tilde r_0^3}{6}- \frac{a^2\tilde r_0}{2}  \right]k^2+ O(k^4)+...,
\end{eqnarray}
Similarly, Eq. (\ref{eq71})  suggest the improved two-parameter expansion
for  $|a|/R > 1$:
\begin{eqnarray}\label{eq811}
 {k}\cot \delta = -\frac{1}{a} +\left[\frac{r_0}{2}-\frac{r_0^3}{6a^2}-\frac{2r_0^2}{\pi^2 a}\right]k^2+O(k^4)+...,
\label{eq711}
\end{eqnarray}
}Expansions (\ref{eq1}) { and (\ref{eq811}) are} appropriate for $|a|/R  > 1$ and expansions (\ref{eq9}) { and  (\ref{eq91})}   for $|a|/R  < 1.$

\begin{figure}[!t]
\begin{center}
\includegraphics[width=\linewidth]{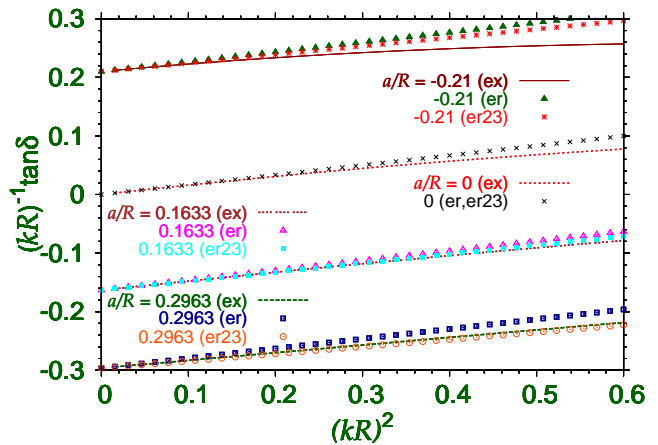} 
\caption{Exact (ex) dimensionless effective-range function  $(kR)^{-1}\tan \delta$ versus dimensionless energy
$(k R)^2$   of   (\ref{eq3}) for different scattering lengths compared with the same from the effective-range 
expansions (\ref{eq9}) (er)  and (\ref{eq91}) (er23) using the effective range $\tilde r_0 =R$.
   }\label{fig2}
\end{center}

\end{figure}

\section{Numerical Result}

To test the performance of the effective-range expansions (\ref{eq9}) and (\ref{eq91})  we consider a square-well potential of range $R$ and depth $\beta^2$. The domain of applicability of this expansion is small energy $k^2$ $ (k^2 \ll \beta^2)$  and small values of scattering length $a$ $ (|a|\ll R)$.  The depth parameter $\beta^2 $ will be conveniently chosen to satisfy the condition $|a|\ll R$.
In  figure \ref{fig1} we plot dimensionless scattering length $a/R$ given by   (\ref{eq4})  versus $\beta R$.
From this figure we find that the scattering length has multiple zeros as a function of $\beta R$. The zero at the origin 
for vanishing strength $\beta R$ of the square well  is trivial. The first nontrivial zero of $a$ appears for  $\beta R \approx 4.4934$; in this case there is a single bound state of the system. 
  We consider four values of $\beta$ near this zero of $a$, e. g., $\beta =4.4,  4.45, 4.4934$ and $4.515$ leading to scattering lengths $a/R=0.2963, 0.1633, 0$ and $ -.21$, respectively, all satisfying  $(|a|/R\ll 1)$. In  figure 2 we plot $(kR)^{-1}\tan \delta$ versus $(kR)^2$ as obtained from the analytic expression (\ref{eq3}) and also from the effective-range expansions (\ref{eq9}) and (\ref{eq91}) taking the effective range equal to the range of the square well:  
$\tilde r_0 =R$.
 From  figure \ref{fig2} we find that  the proposed effective-range expansion (\ref{eq9})   gives a good description of low-energy scattering for small positive and negative values of scattering length ($|a|/R\ll 1$). {Nevertheless, the improved effective-range expansion (\ref{eq91})  gives a better account of the actual state of affairs. }

\begin{figure}[!t]
\begin{center}
\includegraphics[width=\linewidth]{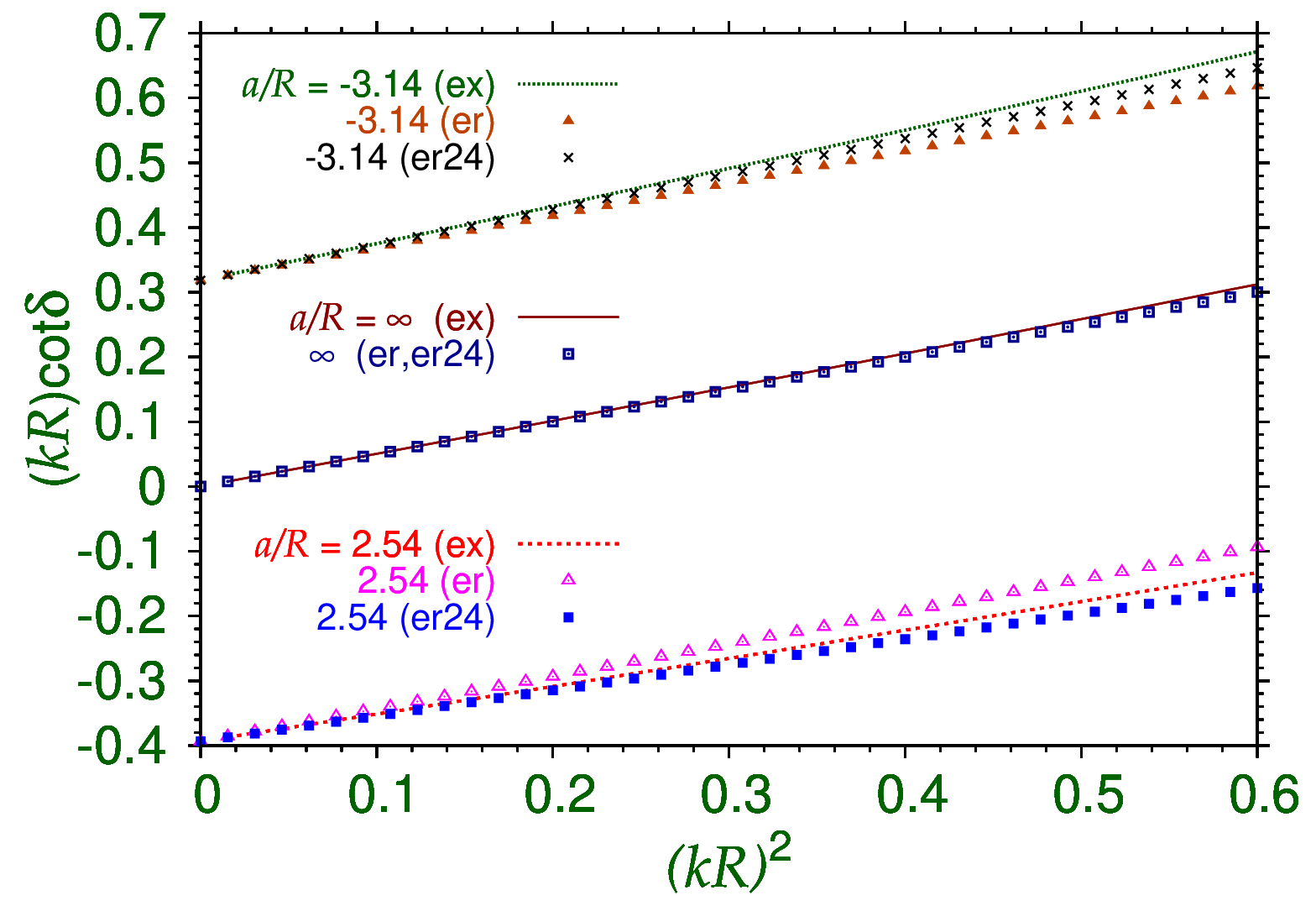} 
\caption{Exact (ex) dimensionless effective-range function  $(kR)\cot \delta$ versus dimensionless energy
$(k R)^2$   of   (\ref{eq1}) for different scattering lengths compared with the same from the effective-range 
expansions (\ref{eq1}) (er)  and (\ref{eq811}) (er24) using the effective range $ r_0 =R$.
   }\label{fig3}
\end{center}

\end{figure}

{
In figure \ref{fig3} we contrast the effective-range expansion (\ref{eq1}) with (\ref{eq811}) by plotting the function $kR\cot \delta $ versus $(kR)^2$ for $a/R= -3.14, 0$ and $2.54$. We display the exact result (\ref{eq3}),  the effective-range expansion (\ref{eq1}), and the improved expansion (\ref{eq811}) in this figure. 
We find that although the expansion (\ref{eq1}) leads to a good description of $kR \cot \delta$ at low energies, the improved expansion (\ref{eq811}) has better performance compared to the exact result. 
}

\begin{figure}[!t]
\begin{center}
\includegraphics[width=\linewidth]{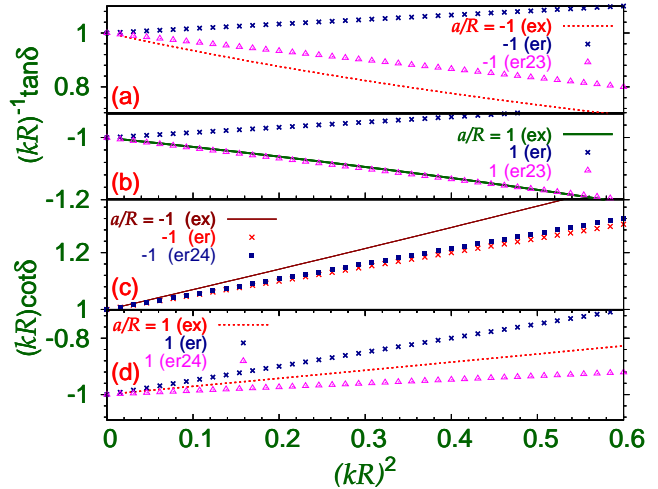} 
\caption{ Exact (ex) dimensionless effective-range function $(kR)^{-1}\tan \delta$ versus dimensionless energy $(k R)^2$  for (a) $a/R =-1$ and (b) $a/R =1$  compared with the same from the effective-range 
expansions (\ref{eq9}) (er)  and (\ref{eq91}) (er23), respectively.  The same for the function 
  $(kR)\cot \delta$ versus dimensionless energy
$(k R)^2$   for  (c) $a/R =-1$ and (d) $a/R =1$
 compared with  the effective-range 
expansions (\ref{eq1}) (er)  and (\ref{eq811}) (er24).
   }\label{fig4}
\end{center}

\end{figure}

{Effective-range expansions (\ref{eq9}) and (\ref{eq91})  on the one hand and (\ref{eq1}) and 
(\ref{eq811}) on the other are valid for $|a|<R$ and $|a|>R$, respectively. It is pertinent to investigate the limits of validity of these expansions.  To  study how well they perform for $|a|\sim R$, we compare   in figure \ref{fig4}(a)-(b)  the results of effective-range expansions  (\ref{eq9}) and (\ref{eq91}) with the corresponding exact results for $a/R=\mp 1$. The same  
 for the expansions   (\ref{eq1}) and (\ref{eq811}) for  $a/R=\mp 1$ are illustrated in figures 
\ref{fig4}(c)-(d). We find that for both   $a/R=\pm 1$  the improved expansions (\ref{eq91})  and (\ref{eq811})  perform better than the basic  expansions  (\ref{eq9}) and  (\ref{eq1}).
Considering that the expansions of this paper are not appropriate for the   conditions   $a/R=\mp 1$   considered in figure \ref{fig4}, these perform fairly well in all cases with the improved expansions (\ref{eq91}) and (\ref{eq811}) exhibiting better performance that the basic expansions (\ref{eq9}) and (\ref{eq1}), respectively.   
}

\section{Summary}

We have proposed an effective-range expansion (\ref{eq9}) valid for   scattering lengths much smaller than the range of the potential: $|a|/R \ll 1$. This should be considered complementary to the textbook expansion (\ref{eq1}) valid for  $|a|/R \gg 1$. Two improved expansions (\ref{eq91}) and (\ref{eq811}) are also suggested for $|a|/R < 1$ and $|a|/R > 1$, respectively.
 We illustrated the usefulness of these expansions with a finite-range square-well potential for both positive and negative scattering lengths. Nevertheless, one should recall that these  expansions  are valid for short-range potentials. In many problems of physical interest the underlying potential is not of short range and such expansions are not physically meaningful. In these cases such an expansion could involve a linear and a logarithmic term in $k$ \cite{xxx}, viz. (3.39) of \cite{thaler}.

\section*{Acknowledgments}  
SKA thanks the Funda\c c\~ao de Amparo \`a Pesquisa do Estado de 
S\~ao Paulo (Brazil) (Projects:  2012/00451-0 and 2016/01343-7) and the Conselho Nacional 
de Desenvolvimento   Cient\'ifico e Tecnol\'ogico (Brazil) 
(Project: 303280/2014-0) for support. 
  

\end{document}